\documentstyle[11pt]{article}
\input psfig
\long\def\comment#1{}
\begin{document}
\title{Representation of Entangled States}
\author{Subhash Kak\\
Department of Electrical \& Computer Engineering\\
Louisiana State University,
Baton Rouge, LA 70803, USA}
\maketitle

\begin{abstract}
Identifying a pair of correlated  
qubits with the pure entangled state $\frac{1}{\sqrt 2} 
( | 00\rangle +  |11\rangle)$ is an idealization unless 
the pair is so prepared using an appropriate quantum gate operating
on a known state.
Questions related to the 
reference frame for measurement of
the entangled state are 
considered.

\end{abstract}

\thispagestyle{empty}

\section*{Introduction}

The only way to certify a specific pure entangled state 
is to start with a known pure state, say $|00\rangle$,
and use an appropriate 2-qubit transformation to drive it
to the desired state.

This note presents some observations concerning
entanglement states.
Specifically, the question of the use of an enlarged
reference frame to test entangled particles is discussed.

\section*{Imperfect entanglement}

In the decay of a spinless system into a pair
of spin $\frac{1}{2}$ particles, say electrons, if the spin of one
electrons is found in a particular orientation to be $\frac{1}{2}$,
then the spin of the other electron is $- \frac{1}{2}$. It is
customary to represent the entangled state as
$ | \psi\rangle = \frac{1}{\sqrt 2} (|\uparrow \downarrow \rangle -
 |\downarrow \uparrow \rangle)$. Likewise,
in atomic SPS cascade, if the two
emitted photons are detected in opposite directions, they
appear to have the same polarization. 
The state of the photons is 
usually represented
by: $ |\psi\rangle = \frac{1}{\sqrt 2} (| 00 \rangle + | 11\rangle)$,
where $0$ and $1$ are horizontal and vertical polarization.

In general, one
seeks states such as
 $ |\psi\rangle = \frac{1}{\sqrt 2} (| 00 \rangle + k | 11\rangle)$,
where $|k| = 1$ in many applications.
Entanglement, where the probability amplitudes
of the terms in the superposition are $1, -1, i, -i$ (i.e. $k=\pm 1, \pm i$)
 may be called 
maximal or perfect.

In experiments on entangled photons
created using spontaneous parametric down-conversion\cite{Ko01,Ko00,Ob00,Ru01},
about one out of 10,000 trials produces an
entangled pair and the probability of getting a double-pair
is even lower\cite{Ko03}. Some sorting procedure is use to post-select
entangled pairs out of the large number
created by the source.

When there is maximal
correlation and $0$ and $1$ are obtained with equal
probability on qubits
tested from the source, the principle of least information requires that we
represent the state by 
a density matrix and not as a pure state.

The degree of entanglement at the point of emission 
depends on the physical process. Some processes may not be perfectly
symmetric with respect to the entangled variable.
The mixedness of an entangled state may require a large ensemble
to establish.

\section*{Generating entangled photons}

We may, in principle, generate
the pure entangled state
$\frac{1}{\sqrt 2} (|00\rangle + |11\rangle)$ by operating on the $| 00\rangle$ 
of a pure state by many operators, such as $U$:

\vspace{0.2in}
 $U = \frac{1}{\sqrt 2}                       \left[ \begin{array}{cccc}
1 & ~0 & 0 & 1 \\
0 & ~1 & 1 & 0 \\
0 & ~1 & -1 & 0 \\
1 & ~0 & 0 & -1 \\
\end{array} \right]$

\vspace{0.2in}

However, no deliberately engineered implementation of 
$U$ can be absolutely precise. Due to the inevitable imprecision
in implementing the components of $U$, the actual
entanglement would not be perfect.

For example, if the implemented $U$ is:

\vspace{0.2in}
 $U' = \frac{1}{\sqrt 2}                       \left[ \begin{array}{cccc}
e^{i \theta_1} & ~0 & 0 & 1 \\
0 & e^{i\theta_2} & 1 & 0 \\
0 & 1 & -e^{i\theta_2} & 0 \\
1 & 0 & 0 & -e^{i\theta_1} \\
\end{array} \right]$
\vspace{0.2in}

\noindent
where $\theta_i s$ are small random errors,
an application of the $4\times 4$ Hadamard operator 

 $H = \frac{1}{2}                       \left[ \begin{array}{cccc}
1 & 1 & 1 & 1 \\
1 & -1 & 1 & -1 \\
1 &  1 & -1 & -1 \\
1 & -1 & -1 & 1 \\
\end{array} \right]$

\vspace{0.2in}

\noindent
on $U' |00\rangle$
will reveal that the
state is not correctly rotated.

\section*{The reference frame}

To test entanglement it is assumed that the experimenters, if they
are at different locations, share the same
reference frame which, in general, will be
three-dimensional.
For photons, the polarization state is determined by
the oscillating electrical or magnetic vectors that are
mutually orthogonal and perpendicular to the direction of
propagation.
The propagation direction is assumed to fix one of the
three axes of the reference frame and, therefore, 
the problem is reduced to one of two dimensions only.

In the general case, one must view a qubit 
$a |0\rangle + b |1\rangle $
as a point on a unit
sphere (Figure 1). Here it is assumed that the measurements are with respect
to the XY-plane; the third axis represents the phase variable $\theta$.

\vspace{2mm}
\begin{figure}
\hspace*{0.5in}\centering{
\psfig{file=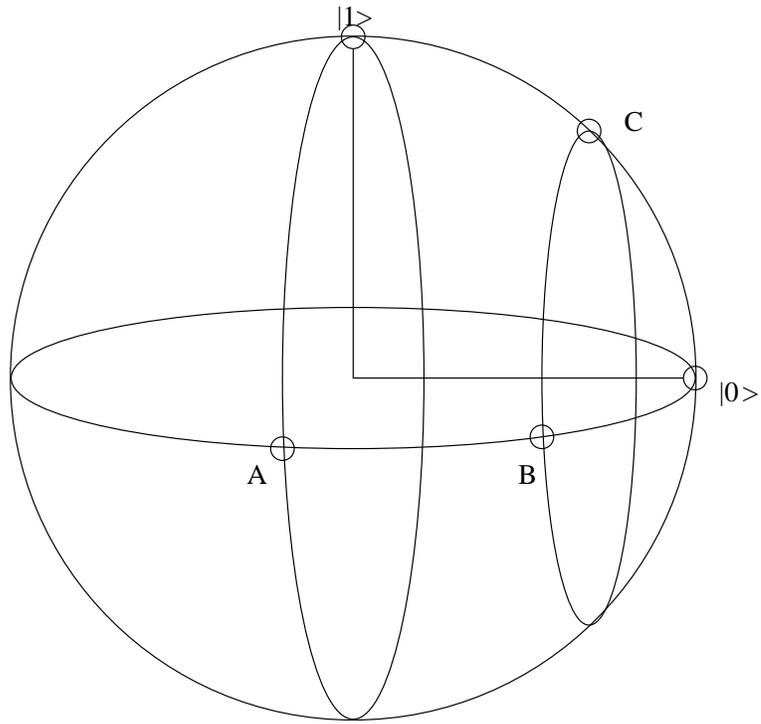,width=10cm}}
\caption{The qubit sphere $( \alpha, \beta, \theta)$.
The vertical circles represent $|1\rangle$ and its phase
shifts.
The circle on the right represents
$ 1/2^{1/2}(|0\rangle + e^{i \theta} |1\rangle ) $, which are
various combinations of $|0\rangle$ with phase shifted $|1\rangle$
(i.e. $45^o$ polarized photons, for example).
The point A is
$  e^{i \pi /2} | 1\rangle$; B is
$ 1/2^{1/2}(|0\rangle + i |1\rangle ) $; C is
$ 1/2^{1/2}(|0\rangle +  |1\rangle ) $.
}
\end{figure}

\vspace{2mm}

The most general rotation operation on the qubit 
$ |0\rangle $ may be represented by
$\left[ \begin{array}{cc}
                                  \alpha e^{i \theta_1}  & \beta e^{-i \theta_2} \\
                                  \beta e^{i \theta_2} & -\alpha e^{-i \theta_1} \\
                               \end{array} \right]$
where $\theta_1$ and $\theta_2$ are phase angles. This leads to the
superposition state
$ \alpha e^{i \theta_1}  | 0\rangle + \beta e^{i \theta_2} | 1\rangle.$

If no reference axis is available then one can use the full 
complement of three axes.
The qubit may then be written as:
$a|0\rangle + b|1\rangle + c|2\rangle $.

Thus in the BB84 quantum cryptography protocol\cite{Be84}, one could 
represent polarization along
9 different directions 
rather than just two.
These 9 directions would be:

\vspace{0.2in}
$ |0\rangle, |1\rangle, |2\rangle, |0\rangle + |1\rangle,
|0\rangle - |1\rangle, |0\rangle + |2\rangle, |0\rangle - |2\rangle,
|1\rangle + |2\rangle, |1\rangle - |2\rangle$

\vspace{0.2in}
These 9 directions belong to 6 different frames in the X, Y, and Z
planes.
Generalized Bell states for such a case can be easily defined\cite{Fu01}.

\subsection*{Enlarged basis set}

In the Bell basis, 
the entanglements are for the
mutually orthogonal states:

\vspace{0.2in}

$|00\rangle + |11\rangle, |00\rangle - |11\rangle,  |01\rangle + |10\rangle, |01
\rangle - |10\rangle$

\vspace{0.2in}
This set may be enlarged by considering weights $\pm i$ and in applications
such as dense coding one may use the basis states containing $\pm i$
rather than the usual Bell basis. The enlarged set of basis states is:

\vspace{0.2in}

$|00\rangle + |11\rangle, |00\rangle - |11\rangle,
  |01\rangle + |10\rangle, |01\rangle - |10\rangle$

$|00\rangle + i |11\rangle, |00\rangle - i |11\rangle,
  |01\rangle + i |10\rangle, |01\rangle - i |10\rangle$

\vspace{0.2in}

The operators given below represent relevant
transformations that form a group:

\vspace{0.2in}
$I =      \left[ \begin{array}{cc}
              1  &  \\
                & 1 \\
               \end{array} \right]$,
$A =      \left[ \begin{array}{cc}
                & 1  \\
              1 &  \\
               \end{array} \right]$,
$B =      \left[ \begin{array}{cc}
              1  &  \\
                & -1 \\
               \end{array} \right]$,
$C =      \left[ \begin{array}{cc}
                & 1  \\
              -1 &  \\
               \end{array} \right]$,

\vspace{0.2in}

$D =      \left[ \begin{array}{cc}
              1  &  \\
                & i \\
               \end{array} \right]$,
$E =      \left[ \begin{array}{cc}
                & 1  \\
              i &  \\
               \end{array} \right]$,
$F =      \left[ \begin{array}{cc}
              1  &  \\
                & -i \\
               \end{array} \right]$,
$G =      \left[ \begin{array}{cc}
                & 1  \\
              -i &  \\
               \end{array} \right]$

\vspace{0.2in}
The multiplications products for the elements of this group are shown
in Table 1.

\vspace{0.2in}

\noindent
{\bf Table 1:} A group of elementary quantum operators where the item
in the left column comes first in the multiplication

\vspace{0.1in}
\begin{tabular}{||c|cccccccc||} \hline \hline
mult  & I & A&B&C&D&E&F&G\\ \hline
I & I & A&B&C&D&E&F&G\\
A & A & I & C & B & G & F & E & D \\
B & B & C & I & A & F & G & D & E \\
C & C & B & A & I & E & D & G & F \\
D & D & E & F & G & B & C & I & A \\
E & E & D & G & F & A & I & C & B \\
F & F & G & D & E & I & A & B & C \\
G & G & F & E & D & C & B & A & I \\ \hline
\end{tabular}

\vspace{0.2in}
If one did not wish to use the operators containing $\pm i$, then the
subgroup consisting of
the operators $I, A, B, C$ will suffice.

Corresponding to the use of the Hadamard operator
$ \frac{1}{\sqrt2}      \left[ \begin{array}{cc}
               1& 1  \\
               1&-1 \\
               \end{array} \right]$ to distinguish between
$|0\rangle + |1\rangle$ and $|0\rangle - |1\rangle$,
one may distinguish between
$|0\rangle + i|1\rangle$ and $|0\rangle - i|1\rangle$
using the operator
$ \frac{1}{\sqrt2}      \left[ \begin{array}{cc}
               1& -i  \\
               i&-1 \\
               \end{array} \right]$.

The idea of the enlarged basis may be used in cryptography\cite{Be84,Be95}.
The four states of the quantum cryptographic protocol may be
viewed as polarizations at -45$^o$, 0$^o$, +45$^o$, +90$^o$ and
these polarizations are recovered by the use of the
0/90$^o$ and -45/+45$^o$ bases. Since the shared reference frame is
0/90$^o$, it makes the two pairs of states asymmetric in the
sense that one pair has no superposition whereas the other does.
The states 
$|0\rangle + |1\rangle,$ $|0\rangle - |1\rangle$
$|0\rangle + i|1\rangle,$ $|0\rangle - i|1\rangle$
constitutes a set where each state is a superposition and it
may be used in place of the usual set.

\section*{Conclusion}

Many applications in quantum information science
require entangled pure states.
In some of those, like dense coding, a small deviation from
maximal entanglement appears to make only a correspondingly
small difference in the final results.
On the other hand, in teleportation one would like to
have a very precisely defined
maximally entangled pure state. 
We could create this entangled state using an appropriate
transformation on a state such as $|00\rangle$.

This indicates the importance of hardware
implementation of basic quantum gates.
This is also significant because quantum lithography\cite{Ko01a}
using
entanglement of groups of photons could
increase etching resolution beyond the diffraction limit.


\end{document}